\documentclass[twocolumn]{aastex62}

\usepackage{graphicx}

\newcommand{\msolar}{M_{\odot}}

\newcommand{{\maxi}}{\rm MAXI}
\newcommand{{\swift}}{\rm Swift}

\graphicspath{{./}{figures/}}

\received{April 7, 2020 }
\revised{July 23, 2020}
\accepted{July 30, 2020}
\submitjournal{ApJL}

\shorttitle{Discovery of MAXI~J1348$-$630}
\shortauthors{Tominaga et al.}

\begin{document}

\title{Discovery of the black hole  X-ray binary transient MAXI~J1348$-$630}

\correspondingauthor{Mayu Tominaga}
\email{tominaga@ac.jaxa.jp}

\author{Mayu Tominaga}
\affiliation{Institute of Space and Astronautical Science (ISAS), Japan Aerospace Exploration Agency (JAXA), 3-1-1 Yoshinodai, Chuo, Sagamihara, Kanagawa, 252-5210, Japan}

\author[0000-0001-9307-046X]{Satoshi Nakahira}
\affiliation{Institute of Space and Astronautical Science (ISAS), Japan Aerospace Exploration Agency (JAXA), 3-1-1 Yoshinodai, Chuo, Sagamihara, Kanagawa, 252-5210, Japan}

\author[0000-0001-8195-6546]{Megumi Shidatsu}
\affil{Department of Physics, Ehime University, 
2-5, Bunkyocho, Matsuyama, Ehime 790-8577, Japan}

\author{Motoki Oeda}
\affil{Department of Physics, Tokyo Institute of Technology, 2-12-1 Ookayama, Meguro-ku, Tokyo 152-8551, Japan}

\author{Ken Ebisawa}
\affil{Institute of Space and Astronautical Science (ISAS), Japan Aerospace Exploration Agency (JAXA), 3-1-1 Yoshinodai, Chuo, Sagamihara, Kanagawa, 252-5210, Japan}

\author{Yasuharu Sugawara}
\affil{Institute of Space and Astronautical Science (ISAS), Japan Aerospace Exploration Agency (JAXA), 3-1-1 Yoshinodai, Chuo, Sagamihara, Kanagawa, 252-5210, Japan}

\author{Hitoshi Negoro}
\affil{Department of Physics, Nihon University, 1-8-14 Kanda-Surugadai, Chiyoda-ku, Tokyo
101-8308, Japan
}

\author{Nubuyuki Kawai}
\affil{Department of Physics, Tokyo Institute of Technology, 2-12-1 Ookayama, Meguro-ku, Tokyo 152-8551, Japan}

\author[0000-0002-1190-0720]{Mutsumi Sugizaki}
\affil{National Astronomical Observatories, Chinese Academy of Sciences, 
20A Datun Rd, Beijing 100012, China}

\author{Yoshihiro Ueda}
\affil{Department of Astronomy, Kyoto University, Kitashirakawa-Oiwake-cho, Sakyo-ku, Kyoto 606-8502, Japan}

\author{Tatehiro Mihara}
\affil{High Energy Astrophysics Laboratory, RIKEN, 2-1 Hirosawa, Wako, Saitama 351-0198, Japan}
 


\begin{abstract}
We report the  first half-year  monitoring of the new Galactic black hole candidate MAXI J1348--630, discovered on 2019 January 26 with the Gas Slit Camera (GSC)
on-board MAXI.
During the monitoring period, the source exhibited two outburst peaks, where
the first peak flux (at $T$=14 day from the discovery of $T$=0) was  $\sim$4 Crab (2--20 keV) and the second one (at $T$=132 day) was $\sim$0.4 Crab (2--20 keV). The source exhibited distinct  spectral transitions between the high/soft and low/hard states and an apparent  ``q''-shape curve on the hardness--intensity diagram,  both of which are well-known characteristics of black hole binaries.
Compared to other bright black hole transients, MAXI J1348--630 is characterized by its low disk-temperature 
($\sim$0.75 keV at the maximum) and high peak flux in the high/soft state.
The low peak-temperature leads to  a large innermost radius that is identified as  the Innermost Stable Circular Orbit (ISCO), determined by the black hole mass and spin.
Assuming the empirical relation between the  
soft-to-hard transition luminosity ($L_{\rm trans}$) and the  Eddington luminosity ($L_{\rm Edd}$),   $L_{\rm trans}/L_{\rm Edd} \approx 0.02$, 
and a face-on disk around a non-spinning black hole,
the source distance and the black hole mass are  estimated to be $D\approx 4$ kpc and   $\sim$7$ \; \left(D / 4\; \rm{kpc} \right) \msolar$, respectively.
The  black hole is    more massive  if the   disk is inclined and the black hole is spinning.
These results suggest that MAXI J1348--630 may host  a relatively massive black hole among the known black hole binaries  in our Galaxy.
\end{abstract}



\section{Introduction} 
Black hole binaries (BHBs), those consisting of a stellar mass black hole and a main-sequence star, are known to exhibit  two distinct X-ray spectral states; the ``low/hard'' state and the ``high/soft'' state \citep[e.g.,][]{2006ARA&A..44...49R,  Belloni2009,mcclintock2009}.
The low/hard state is observed when the mass accretion rate is relatively low, where
the optically-thick/geometry-thin standard  disk is likely to be truncated before reaching the Innermost Stable Circular Orbit (ISCO), and the inner-disk becomes a radiatively inefficient hot accretion flow.
When the accretion rate becomes higher than a certain threshold, the inner-disk switches its nature from the  optically-thin/geometrically-thick state to the standard disk, where the innermost  radius will reach to  the ISCO.

One of the most efficient ways to discover such BHBs and study their state transitions is to continuously monitor the entire sky in X-rays, because most BHBs are transients and  exhibit unpredictable X-ray outbursts.
The Monitor of All-sky X-ray Image ($\maxi$; \citealt{2009PASJ...61..999M}), which is operated on the International Space Station (ISS) and surveying $\sim$85 $\%$ of the sky every $\sim$ 92 minutes (corresponding to the ISS orbital period, where each strip   of the sky is  exposed for  a duration of 40--100 sec), is an ideal instruments for that purpose. 
In fact, since 2009, MAXI discovered 13 new BHB transients and monitored their state transitions extensively \citep{Negoro2019, ATel13256}.

In this letter, we report the discovery of MAXI J1348$-$630 on 2019 January 26 by MAXI \citep{2019ATel12425....1Y} and the results of continuous monitoring of the source till 2019 August 3 (176 days after the discovery) including two outbursts (see also \citealt{2020arXiv200403792J} for the first outburst).
Since its discovery, the source was subsequently observed with other X-ray telescopes, $\swift$/X-ray Telescope \citep{2019ATel12434....1K}, INTEGRAL \citep{2019ATel12441....1A}, and NICER \citep{2019ATel12447....1S}.  We  also analyze  $\swift$  data to supplement the energy bands above $\sim$15  keV and below $\sim$2.0 keV, where  MAXI is not sensitive.

Optical counterpart was detected by iTelescope.Net T31 0.51-m telescope in Siding Spring  \citep{2019ATel12430....1D}, Swift/UVOT   \citep{2019ATel12434....1K}, and Las Cumbres Observatory  network 2-m and 1-m telescopes \citep{2019ATel12439....1R}.
The radio counterpart was detected by Australia Telescope Compact Array \citep{2019ATel12456....1R} with a flat spectrum consistent with a compact jet.

\begin{figure*}[t]
\plotone{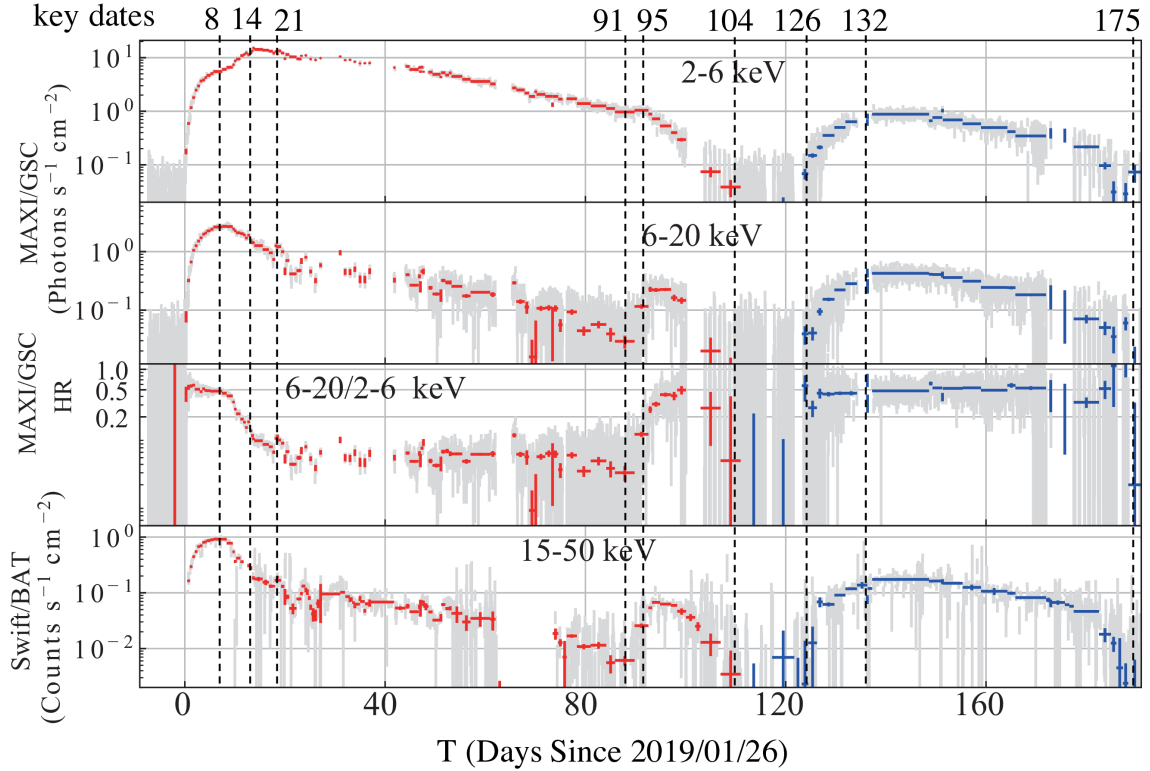}
  \caption{
  The MAXI/GSC 2--6 keV and 6--20 keV light curves, 
  hardness ratio between the two energy bands, and the 
  15--50 keV {\it Swift}/BAT light curve, from top to bottom.
  Grey points indicate individual scans and colored points are adaptively binned data (see the text).
  The difference in color represents the first outburst (red) and the second one (blue).
  Black dashed lines indicate the following key dates; $T=$8 (hard band peak), $T=$14 (soft band peak), $T=$21 (end of the hard-soft transition), $T=$91 (start of soft-hard transition),  $T=$95 (end of the soft-hard transition), $T=$104 (disappearance), $T=$126 (reappearance), $T=$132 (peak after the reappearance), and $T=$175 (second disappearance).
   }
\label{fig:lightcurve}
\end{figure*}

\section{Observation and Data Reduction}
\subsection{MAXI}
MAXI J1348$-$630 was detected with the Gas Slit Camera \citep[GSC;][]{Mihara2011, 2011PASJ...63S.635S}  through the MAXI nova alert system \citep{Negoro2016} at 03:16 UT on 2019 Jan 26 (MJD 58509, $T$=0; hereafter, $T$ is defined as $\rm{MJD}$-58509), where  the X-ray flux was 47$\pm{8}$ mCrab (4--10 keV)   \citep{2019ATel12425....1Y}.  
We analyzed the long term data of MAXI J1348--630 from the discovery to   2019 Aug 3  using the MAXI/GSC on-demand web interface \footnote[1]{\url{http://maxi.riken.jp/mxondem}} \citep{nakahira2013we}. 
The source events were extracted   within a circular region of $2\fdg 1$ radius around  the source position $(RA, Dec)=(207\fdg 053, -63.274)$ \citep{2019ATel12434....1K}.
The background region was  set to a  concentric ring with the outer-radius of $3\fdg 0$.
We used only the  GSC counters with minimum damages, GSC \#2 and \#7 operated at the high voltage of 1550 V, and  \#4 and \#5 at 1660 V. 

\subsection{Swift}
The Burst Alert Telescope (BAT) on-board Swift searches all-sky for Gamma-ray bursts, and also 
provides light curves of major sources  in 15-50~keV. We used the public BAT light curves available on the BAT Transition Monitor website \citep{2013ApJS..209...14K} \footnote[2]{\url{https://swift.gsfc.nasa.gov/results/transients/}}.
We also used the pointing observation  data  of the X-Ray Telescope (XRT)   \citep{Burrows2005}.
We used the on-demand web interface \citep{2009MNRAS.397.1177E}\footnote[3]{\url{https://www.swift.ac.uk/user_objects/}} to reduce  the XRT data, whose observation IDs are 00011107024, 00011107029, and 0001110702.
In the spectral analysis, we adopted the response matrix file $\tt{swxwt0to2s6\_20131212v015.rmf}$ provided via the HEASARC calibration database\footnote[4]{\url{https://heasarc.gsfc.nasa.gov/FTP/caldb/data/swift/xrt/cpf/rmf/}}.

\section{Results} 

\subsection{MAXI/GSC} 
Figure \ref{fig:lightcurve} shows the X-ray light curves of MAXI J1348$-$630 in 2--6 keV (soft) and 6--20 keV (hard) bands, and their  hardness ratio, as well as the light curve in 15--50 keV obtained with Swift/BAT.
Because the Crew Dragon Spacecraft  launched by SpaceX  was located on the line of sight during $T$=36--41, we do not use the data during this period.
%

After the discovery on 2019 January 26 (MJD 58509, $T$=0), the hard band flux rapidly increased and reached the peak on $T$=8. 
The source spectra showed a distinct hard-to-soft transition between $T$=8 and $T$=21 \citep{2019ATel12469....1N, 2019ATel12471....1C, 2019ATel12477....1B}. The source was initially in the hard state, where the hardness ratio was almost constant at around 0.5 until the hard-band peak ($T$=8).
After that, the hard band flux rapidly dropped until $T\sim21$, then slowly declined throughout the following 70 days. 
On the other hand, the soft band flux further increased after the hard band peak, then reached its peak on Feb 9 ($T$=14) at $\sim$1.1$\times$10$^{1}$ photons cm$^{-2}$ s$^{-1}$ (2--6keV).
The flux observed with MAXI/GSC (2--20keV) declined steadily during $T$=21--91 by keeping the constant hardness ratio at around 0.05.
Near the end of the first outburst ($T$=91--95), the source went back to the hard state with the hardness ratio  $\sim$0.5.
After $T$=95, the source gradually faded, and finally reached below the detection limit of MAXI/GSC at $T$=104.

On 2019 May 31 (MJD 58634, $T$=126), the source brightened again and reached the peak flux $\sim$1.0 photons cm$^{-2}$ s$^{-1}$ (2--20 keV)
on $T$=132 \citep{2019ATel12829....1R, 2019ATel12838....1N}, which is about 10 $\%$ of the first peak, followed by steady decline.
The hardness ratio during the re-brightening phase was almost constant at 0.5, thus  the source was in the hard state throughout the second outburst. 
After $T$=175, the source got down below the detection limit again.

\begin{figure}[t]
\plotone{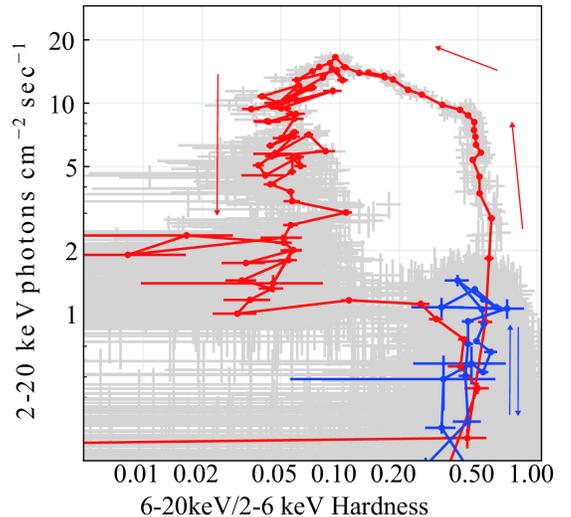}
 \caption{
 Hardness Intensity Diagram (HID) of the hardness ratio (6--20 keV/2--6 keV) vs.\ 2--20 keV intensity. 
  The grey points and the color points connected by line are produced in the same manner as in Figure \ref{fig:lightcurve}.
   Arrows　indicate the  directions along with   dates.}
\label{fig:HID}
\end{figure}

The hardness-intensity diagram (HID) in Figure \ref{fig:HID} shows a clear ``q''-curve \citep[e.g.][]{2019ATel12469....1N}.
According to the time-history of the hardness ratio, we classified the entire observation period into the three spectral states; the low/hard state of $T$=0--8 and 107--191, the transition state of $T$=8--21 and 91--106, and the high/soft state of $T$=21--91.

\begin{figure*}[t]
\plotone{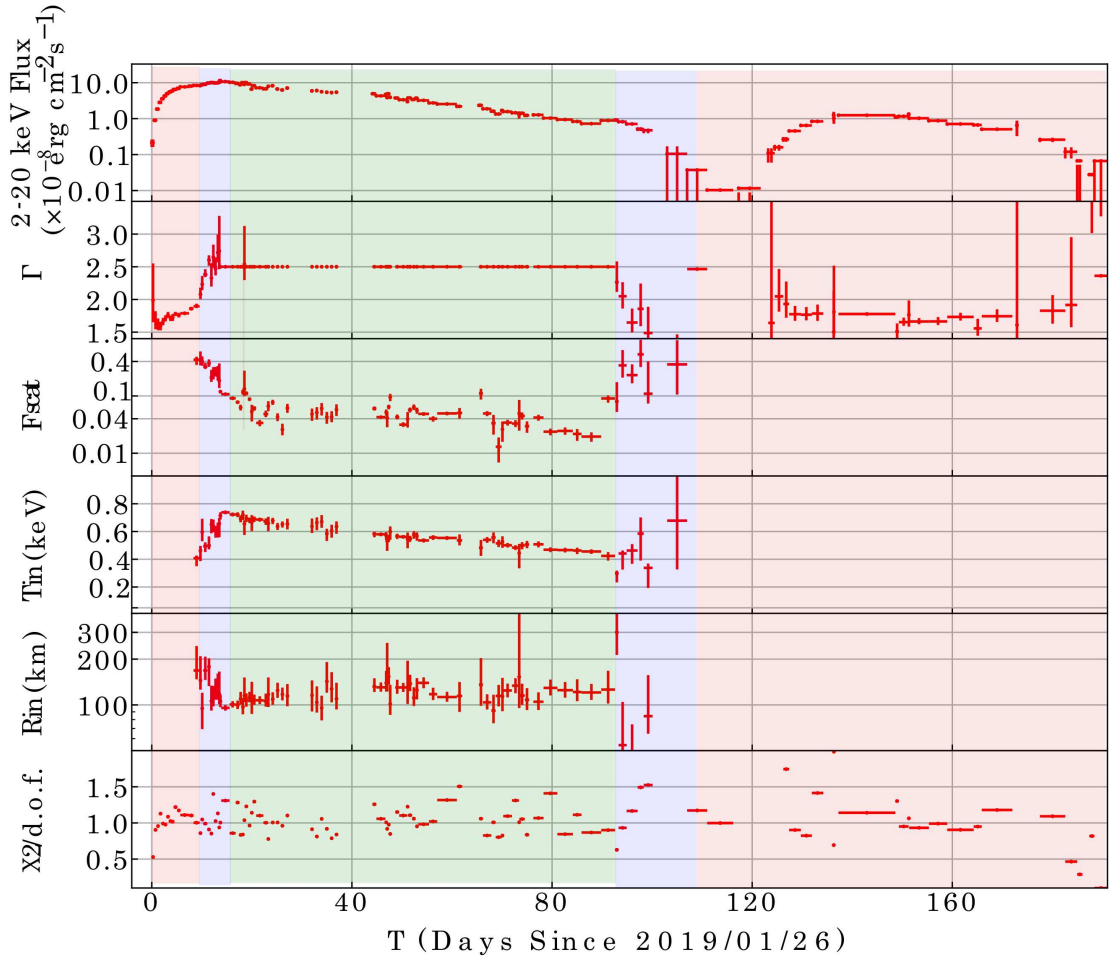}
  \caption{
  Time evolution of the spectral parameters derived from model fits 
  to the 2--20 keV $\maxi$/GSC spectra (90\% confidence).
  $R_{\rm in}$ is  calculated assuming $D$=4 kpc and  $i $=0 (face-on geometry). 
  The background colors represent the spectral states; low/hard state (red; $T=$0--8 and 107--191), transition state (blue; $T=$8--21 and 91--106) and high/soft state (green; $T=$21--91). According to the spectral states, fitting models are different (see texts for details).
   }
\label{fig:spectra}
\end{figure*}
Both the distinct spectral transitions and the ``q''-shaped HID are common properties of BHBs \citep[e.g., ][]{2005Ap&SS.300..107H}. Therefore, we fitted the observed spectra with a standard BHB spectral model.
We used $\tt{xspec}$ version 12.10.0c for spectral model fitting.
Because daily MAXI/GSC data do not have enough photon statistics for model fitting, we adaptively accumulated data for several days using the  Bayesian block representations with the normalization constant $p_0=0.5$ \citep{2013ApJ...764..167S}. 
In the high/soft state, we adopt the optically thick Multi-Color-Disk blackbody model $\tt{diskbb}$ \citep{1984PASJ...36..741M} and its  inverse-Compton scattering approximated by $\tt{simpl}$ \citep{2009PASP..121.1279S}.
We applied the interstellar absorption of Tuebingen-Boulder model $\tt{tbabs}$ with the Solar abundance given by \cite{2000ApJ...542..914W}.
We fixed the neutral hydrogen column density at $N_{\rm{H}} =  8.6\times 10^{21}$ cm$^{-2}$ obtained from Swift/XRT data (see Section \ref{swiftfit}).
The model is described as $\tt{tbabs*simpl*diskbb}$ where free parameters are  scattering fraction $(\it{F_{\rm{scat}}})$ of $\tt{simpl}$, innermost temperature $(\it{T_{\rm{in}}})$ and normalization   ($\it{N_{\mathrm disk}}$) of $\tt{diskbb}$.
The photon index $(\it{\Gamma})$ is assumed to be constant at the canonical value $\it{\Gamma}$=2.5 since the hard-tail in the high/soft state is so weak that the index is hardly constrained \citep[e.g.][]{mcclintock2009}.   Here, we confirmed that the disk temperature $T_{\rm in}$ is only slightly increased or decreased by $\sim$0.03 keV when $\it{\Gamma}$ is changed to 2.0 or 3.0
even when the hard-tail was the strongest ($T$=23).
In the low/hard state, the model is described by $\tt{tbabs*powerlaw}$ where free parameters are  photon index $(\it{\Gamma})$ and  normalization of $\tt{powerlaw}$.
In the transition states, we applied the same model as in the high/soft state, but $\it{\Gamma}$ was treated as a free parameter.
After the phenomenological spectral fittings, we estimate a more realistic  innermost disk radius $R_{\rm in}$ as follows:  Definition of the {\tt diskbb}  normalization $N_{\rm disk}$ is 
\begin{equation}
N_\mathrm{disk} = \left(\frac{r_{\rm{in}}}{D/{10\: \rm{kpc}}}\right)^2 \cos i 
\end{equation}
where $D$ is the source distance,  $i$ is the disk inclination angle, and $r_{\rm in}$ is an apparent  disk radius.
We estimate  $R_{\rm{in}}$ as 
\begin{equation}
R_{\rm{in}}=\xi \kappa^2 r_{\rm{in}}\label{eq:R}
\end{equation}
where $\xi=$0.41 is a correction factor for the inner boundary condition \citep{1998PASJ...50..667K}, $\kappa=$1.7 is the color hardening factor \citep{1995ApJ...445..780S}.
Figure \ref{fig:spectra} shows a time-history of the  spectral fitting parameters (with errors of 90\% confidence limits of statistical uncertainties), where $R_{\rm in}$ is calculated assuming the most plausible distance $D$=4~kpc (see the Discussion below) and $i$=0.  Figure \ref{fig:fitting} shows the results of model fittings on  several key dates.

\subsection{Swift/XRT}\label{swiftfit}
Because the MAXI/GSC data with the energy band above 2 keV cannot precisely determine $N_{\rm{H}}$ of the interstellar absorption,
we  analyzed the data obtained with Swift/XRT, which is more sensitive in  the lower energy range (1--10~keV).
We chose the data of $T$=56, 65, and 85 in the high/soft state, and fitted these spectra with the same  spectral model.
The obtained $N_{\rm{H}}$  value was $\sim$8.6 $ \times 10^{21}$ cm$^{-2}$ at  $T$=85, which we adopt throughout this paper, and   those values in the other dates were agreed within $\sim$2.0$\times 10^{21}$ cm$^{-2}$.

\begin{figure*}[t]
\plotone{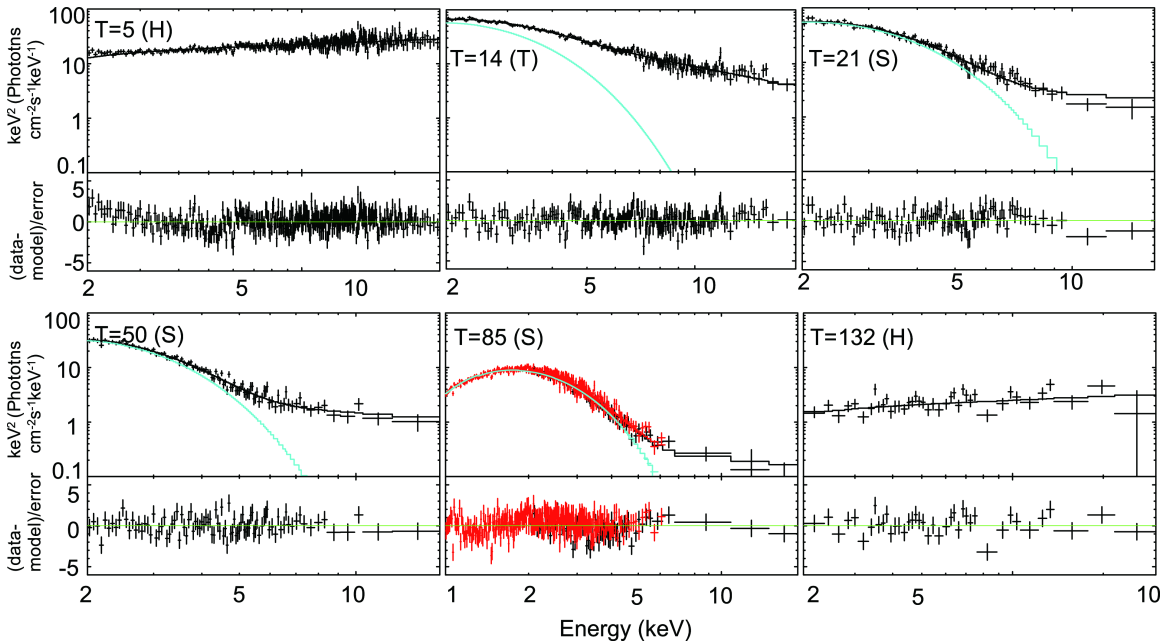}
 \caption{The results of spectral fittings on $T$=5, 14, 21, 50, 85, and 132.
 The alphabets H, T, and S in the brackets mean Hard (low) state, Transition state, and Soft (high) state, respectively.
 The spectra on $T$=85 shows the joint fit of Swift/XRT (red lines) and MAXI/GSC (black lines). 
 For the transition and high/soft state spectra, we plotted the original  $\tt{diskbb}$  spectra  (cyan lines)  without being Compton up-scattering ($F_{\rm scat}$=0) in order to demonstrate the effect of Comptonization.
 }
\label{fig:fitting}
\end{figure*}

Also, we conducted joint spectral fitting of Swift/XRT and MAXI/GSC data taken on $T$=56, 65 and 85. Result on $T$=85 is shown in the bottom-center of Figure~\ref{fig:fitting}. When $T_{\rm{in}}$ was  made independently fitted to each instrument, MAXI GSC gives 0.46 keV, and Swift/XRT gives 0.52 keV.  
We found Swift/XRT always give slightly higher $T_{\rm{in}}$ values than MAXI/GSC during the high/soft state.
This is probably due to  presence of the significant hard-tail, which is hardly constrained by Swif/XRT below $\sim$7 keV but clearly recognized by MAXI.

\section{Discussion} 

We discuss nature of the new X-ray transient MAXI J1348--630 from the results of the data analysis above.
The source exhibited clear spectral transitions
(Figure \ref{fig:lightcurve}) and a q-shaped track on the HID (Figure \ref{fig:HID}), both of which are well-known characteristics of BHBs. Furthermore,
we successfully fitted energy spectra in the three spectral states with their typical spectral models. (Figure \ref{fig:spectra}).  These facts strongly suggest
that MAXI J1348--630 is a new BHB.

Here, we point out that the maximum disk temperature is as low as $T_{\rm in} \approx$ 0.75~keV even at the high/soft state luminosity peak ($T$=14).
This is remarkably lower compared to other luminous black hole transients, where the maximum disk temperature almost always exceeds $\sim$1~keV.
The low disk temperature and the high luminosity lead to the large  innermost radius because the disk luminosity
is proportional to $R_{\rm in}^2 \, T_{\rm in}^4$.
The large innermost radius suggests that MAXI J1348--630 harbors a relatively massive black hole compared to other black hole.

In particular, the innermost radius of the accretion disk ($R_{\rm in}$) is nearly constant  during the high/soft state,  while the 2--20 keV flux and the disk temperature were significantly variable  (Figure \ref{fig:spectra}). This is  a remarkable property in the high/soft state BHBs, such that the constant radius corresponds to the Innermost Stable Circular Orbit \citep[ISCO; e.g.,][]{1989ESASP.296....3T,1993ApJ...403..684E,2010ApJ...718L.117S}, which is determined by  black hole mass and spin.
\begin{figure}[t]
\plotone{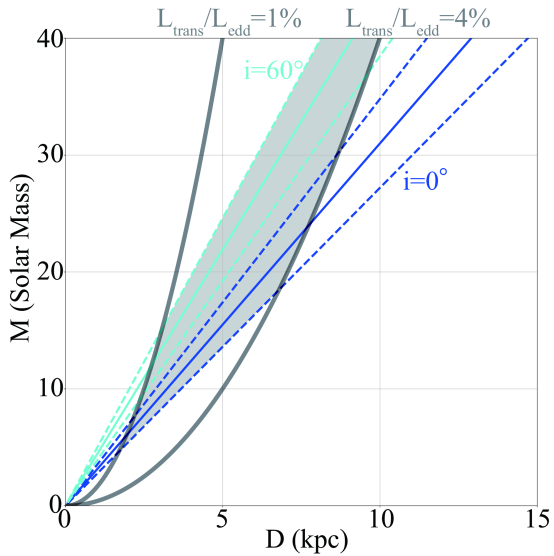}
 \caption{Observational constrains on the distance-mass diagram. Two solid colored lines represent    Equation \ref{eq:mass} when $i=0^{\circ}$ (blue) and $60^{\circ}$ (cyan). 
 The dashed lines with the same colors denote the error region due to systematic error.
 The grey curves give the empirical relation that the luminosity after the soft-to-hard transition ($T=$96) is 0.01--0.04 $  L_{\rm{Edd}}$.
 The shaded region represents the  plausible mass and distance range.
 }
\label{fig:mass}
\end{figure}
In the case of non-rotating black holes, ISCO is equal to three times the Schwarzschild radius. Thus,
from the average innermost radius of the disk, $R_{\rm in} \approx 114 \pm14  \:(D/4 \; {\rm kpc}) (\cos i )^{-1/2}$ km, the  black hole mass $M_{\rm{BH}}$ is estimated to be
\begin{equation}
M_{\rm{BH}}=\frac{c^2R_{\rm{in}}}{6G} \approx 13 \pm 2  \left(\frac{D}{4\; \rm{kpc}}\right)(\cos i)^{-\frac{1}{2}}\msolar. \label{eq:mass}
\end{equation}
The expected distance-mass relation in $i=0^\circ$ and $60^\circ$ are shown in Figure \ref{fig:mass}.

In addition, BHBs  are known to show similar luminosity
dependence of the spectral states.
For instance, in the current case of MAXI J1348--630, the flux at the soft-to-hard transition ($T$=91) is $\sim$10 \% 
of the peak flux in the high/soft state ($T$=14); this value is consistent with those of other  BHBs discovered by \maxi, such as 
MAXI J1820+070 ($\sim$12 \%; \citealt{2019ApJ...874..183S}) and
MAXI J1910-057 ($\sim$10--15 \%; \citealt{2014PASJ...66...84N}).
Furthermore, it was pointed out that the soft-to-hard transition typically occurs at 1--4\% of the  
Eddington luminosity ($L_{\rm{Edd}}$) \citep[]{2003A&A...409..697M}. Namely,
 the bolometric luminosity soon after the soft-to-hard transition is expected to be 0.01 to 0.04 $L_{\rm{Edd}}$.  
The grey curves in Figure \ref{fig:mass} give the distance-mass relation of $L_{\rm{trans}}/L_{\rm{Edd}}= $ 0.01 and 0.04, where we employed the bolometric flux at the soft-to-hard transition $\sim 1.7 \times 10^{-8}$ erg s$^{-1}$ cm$^{-2}$ on $T=$96 and  $L_{\rm{Edd}}=1.3 \times 10^{38} (M/M_\odot)$ erg s$^{-1}$.


Next, we try to constrain the source distance.
The galactic coordinate of the source is $(l,b)=(309.3, -1.1)$, which is tangential to  the Galactic Scutum-Centaurus arm.
If MAXI J1348$-$630 locates in the Scutum-Centaurus arm, the distance is estimated to be $\sim$4--8~kpc \citep[e.g.,][]{2018RAA....18..146X}.
However, the value of  $N_H = 8.6 \times 10^{21}$ cm$^{-2}$ implies that the source is located in front of the arm at $\sim$3--4~kpc.
If we adopt the most likely  index $L_{\rm{trans}}/L_{\rm{Edd}} \approx $ 0.02 \citep{2019MNRAS.485.2744V},  the  distance becomes   $\sim$3.8~kpc for the face-on disk. 
It is  consistent with the estimates from optical observations \citep{2019ATel12439....1R, 2019ATel12480....1C}.

In order to estimate the black hole mass more precisely taking account of the relativistic effects, we tried the $\tt{kerrbb}$ model insted of the $\tt{diskbb}$ model\citep{2005ApJS..157..335L}. 
The $\tt{kerrbb}$ model  gives   relations among the spinning parameter $a$, inclination angle $i$ and black hole mass $M$ under a given source distance, mass accretion rate, and spectral hardening factor ($\kappa$ in Equation \ref{eq:R}).
We carried out a  joint fit to the MAXI/GSC and Swift/XRT spectra at $T$=85 fixing   $D$=4~kpc and $\kappa=1.7$
 for several discrete $a$ and $i$ values, only allowing  the black hole mass and mass accretion rate to be free.
Consequently, the values of $M/M_\odot$ were   7.0 ($a$=0, $i$=$0^\circ$), 14 ($a$=0, $i$=$60^\circ$), 18 ($a =0.998$, $i$=$0^\circ$), and 76 ($a =0.998$, $i$=$60^\circ$). Hence,  we have reached a robust conclusion that MAX J1348--630 hosts a relatively massive black hole.
\cite{2020arXiv200403792J} also suggests a black hole mass of $\sim$9 $M_\odot$ based on an independent accretion disk spectral model.



The binarity of MAXI J1348--63 has  never been confirmed from observations of the  dynamical motion. 
Follow-up detailed optical/infrared spectroscopic observations are strongly encouraged to constrain the black hole mass as well as  the source distance more precisely.

This research has made use of the MAXI data provided by RIKEN, JAXA and the MAXI team, and the Swift data and analysis software provided by the High Energy Astrophysics Science Archive Research Center (HEASARC), which is a service of the Astrophysics Science Division at NASA/GSFC.
        
\bibliographystyle{aasjournal}
\bibliography{ms}

\end{document}